\documentclass[preprint,12pt]{elsarticle}

\usepackage{amssymb}
\usepackage{physics}
\usepackage{float}
\usepackage{graphicx}
\usepackage{sidecap}
\usepackage{tabularx}
\usepackage{array}
\usepackage{booktabs}
\usepackage[dvipsnames]{xcolor}
\usepackage{tikz}
\usepackage{longtable}
\usepackage{hyperref}

\makeatletter
\def\ps@pprintTitle{%
  \let\@oddhead\@empty
  \let\@evenhead\@empty
  \let\@oddfoot\@empty
  \let\@evenfoot\@oddfoot
}
\makeatother

\begin{document}

\begin{frontmatter}

\title{Psitrum: An Open Source Simulator for Universal Quantum Computers}

\date{} 


\author[inst1,inst2]{Mohammed Alghadeer\corref{cor1}}

\affiliation[inst1]{organization={Department of Physics, Clarendon Laboratory, University of Oxford},
            city={Oxford},
            postcode={OX1 3PU}, 
            country={United Kingdom}}
            
\affiliation[inst2]{organization={Department of Electrical Engineering, King Fahd University of Petroleum and Minerals},
            city={Dhahran},
            postcode={31261}, 
            country={ Saudi Arabia}}
            
\cortext[cor1]{mohammed.alghadeer@physics.ox.ac.uk}

\author[inst3]{Eid Aldawsari}

\affiliation[inst3]{organization={Department of Information and Computer Science, King Fahd University of Petroleum and Minerals},
            city={Dhahran},
            postcode={31261}, 
            country={ Saudi Arabia}}
            
\author[inst4]{Raja Selvarajan}

\affiliation[inst4]{organization={Department of Chemistry, Department of Physics and Astronomy, and Purdue Quantum Science and Engineering 
Institute, Purdue University},
            city={West Lafayette, IN},
            postcode={47907}, 
            country={USA}}
            
\author[inst3]{Khaled Alutaibi}

\author[inst4]{Sabre Kais}

\author[inst2]{Fahhad H Alharbi}

\begin{abstract}
Quantum computing is a radical new paradigm for a technology that is capable to revolutionize information processing. Simulators of universal quantum computer are important for understanding the basic principles and operations of the current noisy intermediate-scale quantum (NISQ) processors, and for building in future fault-tolerant quantum computers. In this work, we present simulation of universal quantum computers by introducing Psitrum -- a universal gate-model quantum computer simulator implemented on classical hardware. The simulator allows to emulate and debug quantum algorithms in form of quantum circuits for many applications with the choice of adding variety of noise modules to simulate decoherence in quantum circuits. Psitrum allows to simulate all basic quantum operations and provides variety of visualization tools. The simulator allows to trace out all possible quantum states at each stage M of an N-qubit implemented quantum circuit. Psitrum software and source codes are freely available at: {\color{Blue} \href{https://github.com/moghadeer/Psitrum}{https://github.com/MoGhadeer/Psitrum}}.

\end{abstract}


\end{frontmatter}

\section{Introduction}

Quantum computation is a radical new candidate for a technology that is capable to make a paradigm shift in information processing \cite{In1}. Quantum computers are now a reality with available quantum testbeds and variety of quantum algorithms \cite{In2}. Computation based on quantum algorithms have proved to be more efficient in processing information and solving wide range of complex problems \cite{In3, In4}. Quantum simulators can be designed using quantum algorithms represented by quantum circuits and based on mathematical unitary operations \cite{In5}. In this sense, simulators are special purpose quantum circuits designed to provide insight about specific physical problems. Variety of universal logic gates based on quantum mechanics can be combined to provide very powerful computing features \cite{In6}. Such simulators permit understanding quantum systems that are challenging to study in laboratories and impossible to model with the most powerful supercomputers \cite{In7, In8}.

Universal quantum simulators are based on a quantum computer proposed by Yuri Manin in 1980 \cite{In9} and Richard Feynman in 1982 \cite{In10}. Feynman showed that a Classical Turing Machine (CTM) would not be able to simulate complex quantum systems because of the huge amount of information contained with exponential increase in the required size of computational bits, while his hypothetical universal quantum computer can mimic many quantum effects, such as superposition and entanglement, which allows to efficiently simulate quantum systems \cite{In11}. Quantum properties have shown to significantly enhance simulation power and computational speed when implemented in quantum algorithms \cite{In12, In13}. 

Quantum simulators may be constructed with generally programmable quantum computers \cite{In14}, which would be capable for solving a wider class of problems using both classical and quantum algorithms. The later can be defined as a finite sequence of steps for solving a problem in which each step can be executed on a quantum computer.  \cite{In15, In16}. Instead, a quantum computer simulator can be implemented on the currently available classical hardware to mimic the operation of a real quantum computer, but with limited number of quantum bits (qubits) \cite{In17}. Such simulator allows to test and implement quantum algorithms for many applications even before a fully-controllable quantum computer exists \cite{In18}. Quantum simulators are not to be confused with simulations of quantum
computation on classical hardware which we call a quantum computer simulator.

A universal simulator of quantum circuits can be implemented based on David Deutsch model for a Quantum Turing Machine (QTM) \cite{In19}, which provides a simple model that captures all the power of quantum computation. Therefore, any quantum algorithm can be expressed formally as a particular QTM. The practical equivalent model is a quantum circuit defined as a quantum algorithm implemented on a gate-model based quantum computer with special logic gates. In such circuits, only matrix multiplication and tensor products are the advanced mathematical operations that are used. QTM can be related to classical and probabilistic Turing machines in a framework based on transition matrices. As shown by Lance Fortnow \cite{In20}, a matrix can be specified whose product with the matrix representing a classical or probabilistic machine provides the quantum probability matrix describing the quantum circuit.

Most of the existing quantum computers are still noisy and located in research labs \cite{In21}. In addition, the hardware and maintenance of such systems are expensive and not optimized yet \cite{In22}. Hence, only limited public access to these computers is available. On the other hand, quantum computer (QC) simulators can be useful to overcome such problems. A QC simulator is a software program which imitates the functionality of a quantum computer using classical hardware \cite{In23}, which allows to design, run and test quantum algorithms. There are many QC simulators available to the public \cite{In24}. They differ in purpose, language, size, complexity, performance and technical-based type (e.g. toolkits) \cite{In25}. Companies working in this field (e.g. IBM, Microsoft, Rigetti, Google, and ETH Zurich) are creating full-stack libraries for universal quantum computers with useful simulation tools \cite{In25, In26, In27}. Most of these tools are not software, but software development kits (SDKs) or frameworks, including Qiskit \cite{In28}, LIQUi$\ket{}$ \cite{In29}, ProjectQ \cite{In30}, Cirq \cite{In31}, QX \cite{In31b} and Quantumsim simulator \cite{In31c}. Other simulators focus in enhancing some aspects, such as qHiPSTER from Intel \cite{In32} that takes maximum advantage of multi-core and multi-nodes architectures, and QuEST from Oxford \cite{In33} which is a multithreaded, distributed and GPU-accelerated. There are few software that allow the user to graphically design a quantum circuit and test it without writing a programming code and most of them are web-based, such as IBM Quantum \cite{In34} and Quirk \cite{In35}.

Many QC simulators focus on increasing the number of qubits and simulation speed by using different software tools in order to enable simulating real world quantum algorithms and use cases that can benefit from quantum computing. For example, Quantumsim implemented several optimizations to enable a full density matrix simulation of surface codes in order to evaluate its resilience to noise and whether it can benefit current quantum computing systems to achieve fault-tolerant quantum computation \cite{In31c}. The challenge here is the exponential increase in the dimension of available quantum states with number of qubits \cite{In37,In38}. Furthermore, there are serious shortcomings in the development of abstractions and visualizations of the QC simulation problem. One of these problems is tracing the state of each qubit after each unitary quantum operation \cite{In39}. Another shortcoming is in providing useful probabilistic visualizations of the resulted quantum states \cite{In40}. In this work, we address these issues with many visualizations tools. There are other shortcomings that need to be resolved later, such as using different simulation methods for visualizing quantum states, such as using Feynman path integral formulation \cite{In41} and using tensor networks \cite{In42}.

Quantum computer simulators are important for understanding the operation of noisy intermediate-scale quantum (NISQ) processors, and for building future quantum computers. Current and near-term NISQ computers are limited by the presence of different types of quantum noise that are still too large to allow solving relevant scientific problems. One main source of noise is readout errors that occur during measurements \cite{In43, In44}. They typically prevent reading the correct state of qubits, such as reading zero while the correct state is one and vice versa. Another important source of quantum noise is gate errors, which can be classified into coherent and incoherent noise \cite{In45, In46}. Coherent noise preserve state purity of quantum systems and result in miscalibration in control parameters \cite{In47}. Incoherent noise can be modeled as coherent noise with stochastic varying control parameters. This allows to convert coherent errors into incoherent errors through randomized compiling \cite{In48, In49, In50}. Incoherent noise are relatively easier to handle because they can be modeled as a process that entangles the quantum system with its environment. One very important class of Incoherent noise includes depolarizing channels.

In this work, we present Psitrum -- a universal gate-model based quantum computer simulator implemented on classical hardware. The simulator allows to emulate and debug quantum algorithms in form of quantum circuits for many applications with the choice of adding quantum noise that limit coherence of quantum circuits. Psitrum allows to keep track of quantum operations and provides variety of visualisation tools. The simulator allows to trace out all possible quantum states at each stage M of an N-qubit implemented quantum circuit. The design of Psitrum is flexible and allows the user later to add more quantum gates and variety of noise modules.

\section{Software Structure}
In this work, we use MATLAB to build Psitrum based on universal quantum gates. A set of single- and multi-qubits gates is defined as the basic building blocks of Psitrum. Any other arbitrary multi-qubits operation can be simply implemented by applying successive tensor products and multiplications of the basic gates. It is sufficient to implement these specific gates as a sequence of arithmetic operations on the input state vector in order to implement any quantum logic operation \cite{ST1}. The definition of the basic set of quantum gates is presented in section of additional information below. In addtion, MATLAB is a well-known software environment that can handle and manipulate matrices in form of unitary operations very efficiently \cite{In36}. Psitrum calculates the resulting circuit matrix of any quantum circuit and, in parallel, the software calculates the density matrix as well for applying noise models on the simulation problem. Both the circuit and density matrices are usually very Sparse matrices and, for large circuits, the Sparse packages provided by MATALB makes the simulation problem very efficient in which it subsequently reduce the memory size requirement and speedup the simulation up to a reasonable number of qubits. While this software was built using MATALB, Psitrum is a toolbox that can also be used externally in other platforms, such as python, by using MATLAB APIs.

\subsection{Basic operation}
For a given quantum circuit and an initial vector state, Psitrum calculates the algorithm matrix, the density matrix and the output quantum states which are used to provide useful visualizations. The simulation process of Psitrum starts by replacing each quantum gate by its unitary matrix. Next, the different M stages are combined by applying N-1 tensor products and M-1 matrix multiplications. Then, the output vector state of N qubits is calculated by multiplying the algorithm matrix by the initialized input vector state of the qubits.

The state of an elementary storage unit of a quantum computer is a qubit, described by a two-dimensional vector of Euclidean length one. The normalized state $\ket{\Phi}$ of a qubit can be written as a linear superposition of two orthogonal basis $\ket{0}$ and $\ket{1}$:

\begin{equation}
    \label{eq1}
 \ket{\Phi} \,= a_{0}\ket{0} + a_{1}\ket{1} 
\end{equation}

where $a_{0}$ and $a_{1}$ are complex numbers that satisfy the normalization condition $\abs{a_{0}}^{2} + \abs{a_{1}}^{2} = 1$. In general, the normalized state $\ket{\Psi}$ of N qubits is accordingly described by $2^{N}$ dimensional unit vector:

\begin{equation}
    \label{eq2}
 \ket{\Psi} \,= a_{(0...00)}\ket{0...00} + a_{(0...01)}\ket{0...01} + ...                        a_{(1...10)}\ket{1...10} + a_{(1...11)}\ket{1...11}
\end{equation}

where now the complex coefficients must satisfy the normalization condition:

\begin{equation}
    \label{eq3}
  \sum_{n=0}^{2^{N}-1} \abs{a_{n}}^{2} = 1 
\end{equation}

where $a_{0} = a_{(0...00)}$, $a_{1} = a_{(0...01)}$....and $a_{2^{N}-1} = a_{(1...11)}$. In order to satisfy this condition in Eq. \ref{eq3}, the complex-valued amplitudes $a_{n}$ are rescaled such that $\bra{\Psi}\ket{\Psi} = 1$. State representation of qubits in Psitrum follows the convention in quantum computing literature where the qubits are labeled from $0$ to $N-1$, that is the rightmost (leftmost) bit corresponds to the $0$ ($N - 1$) qubit \cite{ST2}.

\subsection{Framework of Psitrum}
\begin{figure*} [htp]
\includegraphics[width=1.0\textwidth]{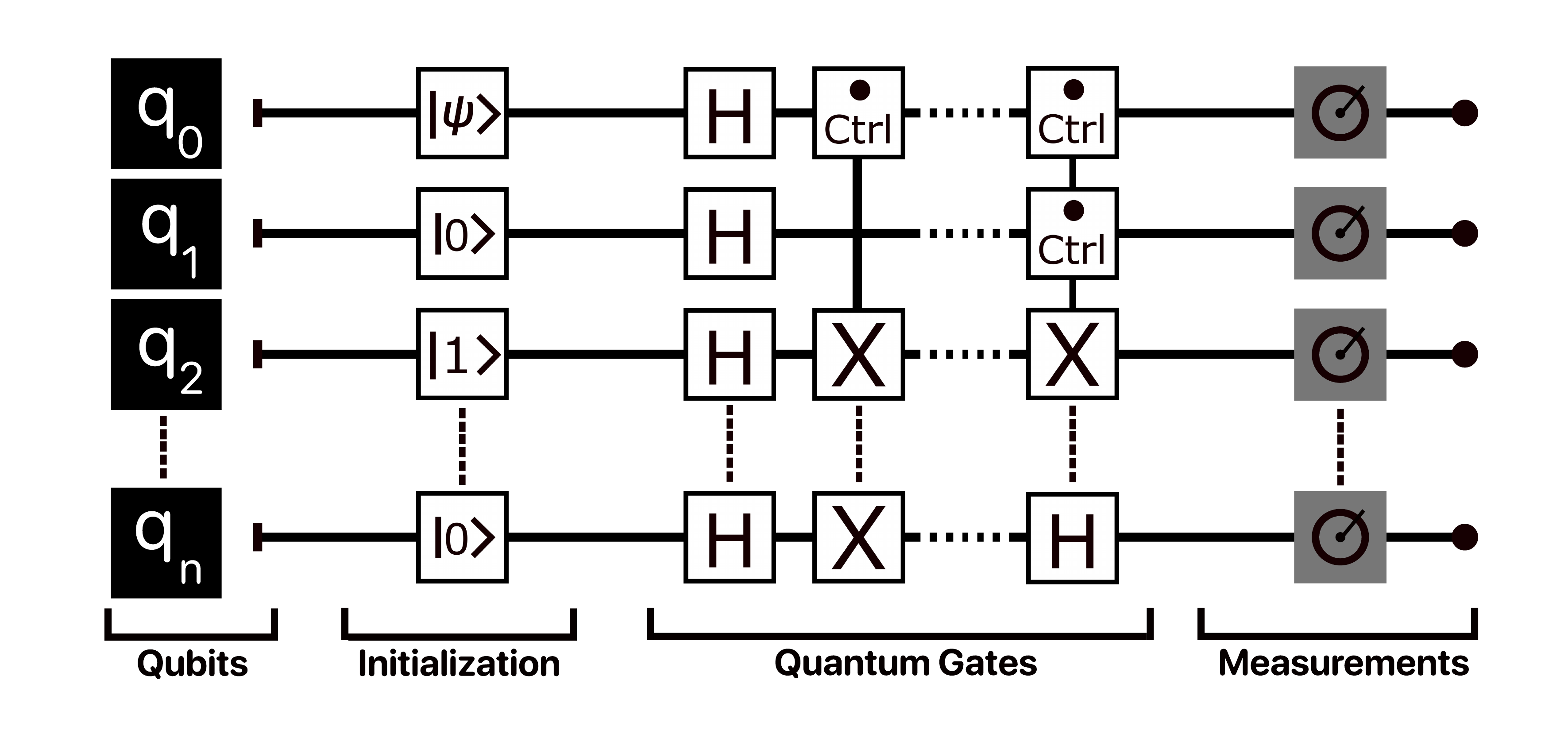}
\caption{Different level of a basic quantum circuit implemented in Psitrum. This framework allows to set/reset qubits, set parameters of quantum gates and choose which output qubits to be measured.}
\label{F1}
\end{figure*}

Quantum algorithms can be fed into Psitrum as string matrices in form of quantum circuits. Rows of the matrices represent N qubits, and columns represent the execution of M operations of the input algorithms. Each element of the circuit represent a gate that applies a specific unitary operation. The definition of each level of a circuit implemented in Psitrum is shown in Fig. \ref{F1}. This framework allows the user to initialize qubits, set parameters of quantum gates and select the output qubits to be measured.  

\begin{figure*} [htp]
\includegraphics[width=1.0\textwidth]{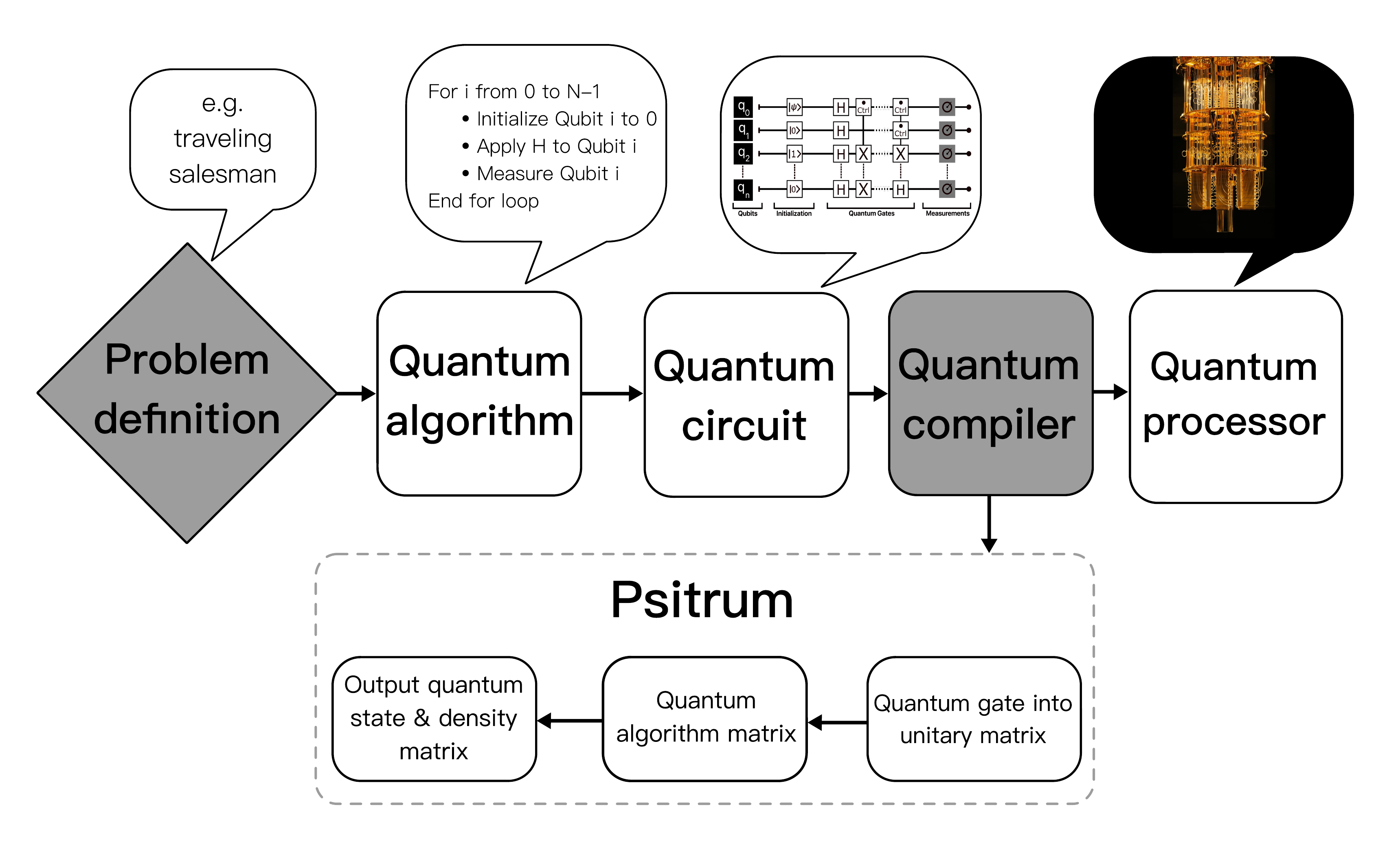}
\caption{Workflow of Psitrum, which follows a full-stack gate-model quantum computer simulator \cite{ST6}. In this way, the selected quantum algorithm for a specific problem needs to be expressed in form of a circuit using a set of defined gates. Psitrum then simulates the operation of an actual quantum compiler that executes this circuit. Finally, a set of visualization tools is available for the user, including Bloch sphere diagrams, state and density tracers, 3D and heatmap diagrams.}
\label{F2}
\end{figure*}

Psitrum is a gate-model quantum computer simulator that follows the workflow of many full-stack quantum simulators \cite{ST3, ST4, ST5, ST6}. First, the problem is defined at a high-level of abstraction, and based on the type of the problem a quantum algorithm is selected in such a way that maximize the output probabilities of the solution. Next, the quantum algorithm is expressed in form of a quantum circuit that applies unitary operations. This circuits then needs to be compiled to a specific set of quantum gates. Finally, the circuit is executed on Psitrum which in turn acts like a quantum compiler. Fig. \ref{F2} shows the workflow of Psitrum with examples and details at each level.  

\subsection{Decoherence and quantum noise}
Simulating quantum errors in Psitrum is possible by simply adding the unitary operation corresponding to a specific noise model. This can be done in the backend of Psitrum. Here, we show how to do that by implementing depolarizing channel which is an important type of incoherent noise. The depolarizing noise model is given by \cite{ST7}:

\begin{equation}
    \label{eq4}
 \xi(\rho) \,= (1-p)\rho+\frac{pI}{2^n}
\end{equation}

where $\xi$ denotes the depolarizing noise channel, $\rho$ is the density matrix, $n$ is the number of qubits and $p$ is the error rate in range $0\leq p \leq \frac{4^n}{4^n -1}$. The parameter $p$ represents the probability that the qubits are depolarized and replaced with completely mixed state $\frac{I}{2^n}$, while $1-p$ is the probability that the qubits are still in their pure states $\rho$. Psitrum gives the choice to select $p$ as a stochastic error rate or at overshoot. The effect of the depolarizing channel is basically a uniform contraction of all axes on the Bloch sphere as function of $p$.

\section{User Interface and visualization}
There are few user friendly and easy-accessible software that allow to graphically design, run, and test quantum algorithms as most of the available QC simulators are SDKs or frameworks. Also, there are even fewer software that allow to represent the simulation results with variety of visualization tools. Psitrum is a software that provides all of these services in an integrated environment. Psitrum provides a simple, friendly, and dynamic graphical user interface (GUI), which is divided into four sections that provide the main services of the software, including circuit designer, quantum state tracer, visualization graphs and numerical representations.

\subsection{Circuit designer}
This section allows the user to graphically design quantum circuits with a defined universal set of quantum gates. The user can add qubits and stages to the circuit as many as needed, and limited only by the available computational power on the user's hardware. The user can then set the initial state of each qubit and which of the output qubits to be measured. Also, it is possible to add additional parameters to some quantum gates, like phase angles of rotations on Bloch sphere. Finally, when the user has designed circuit, next it can be executed on the simulator which dynamically presents the outputs in multiple result windows.

\subsection{Visualization diagrams}
Quantum computers are probabilistic systems and the output of a quantum circuit is based on probabilities. Diagrams can be useful to visualize probabilities, heatmap and Bloch Spheres. These diagrams are provided in Psitrum to visualize algorithm matrix, output states and density matrices.

\subsection{State and density tracers}
The implementation of the quantum state and density matrix tracers in the GUI are another useful features provided by Psitrum. These allow the user to trace out the quantum state of all qubit and desity matrix at each stage after any quantum operation. Psitrum visualizes each step on N qubit in N different Bloch Sphere diagrams to show quantum states at every stage of the circuit. Each qubit is visualized by calculating its Bloch vector. The user can trace out the evolution of all qubits after each stage, either by choosing to access the Bloch vector of each stage or the output Bloch vectors after executing all operations. Similarly, Psitrum shows all density matrices after M stages with and without qunatum noise.

\subsection{Numerical tables}
It is useful, especially for small circuits, to track the state of qubits during the simulation. The ability to observe the numeric values of the output is useful for large circuits where visualization tools are now more needed. In addition, Psitrum also provides several output numerical tables that can be exported, including the output table of the algorithm matrix, its density matrix and the output quantum states. The user can then save the data and have a complete reference of the solution to the simulated quantum algorithm.

\section{Testing and Validation}
This section is about testing Psitrum to validate its performance by implementing four quantum algorithms, quantum full-adder \cite{TV1}, Deutsch-Joza \cite{TV2}, Grover Search \cite{TV3} and prime factorization \cite{TV4} algorithms. These circuits are good benchmark problems for a universal quantum computer simulator \cite{TV5, TV6, TV7, TV8}.

\subsection{Quantum full-adder}
\begin{figure*} [htp]
\includegraphics[width=1.0\textwidth]{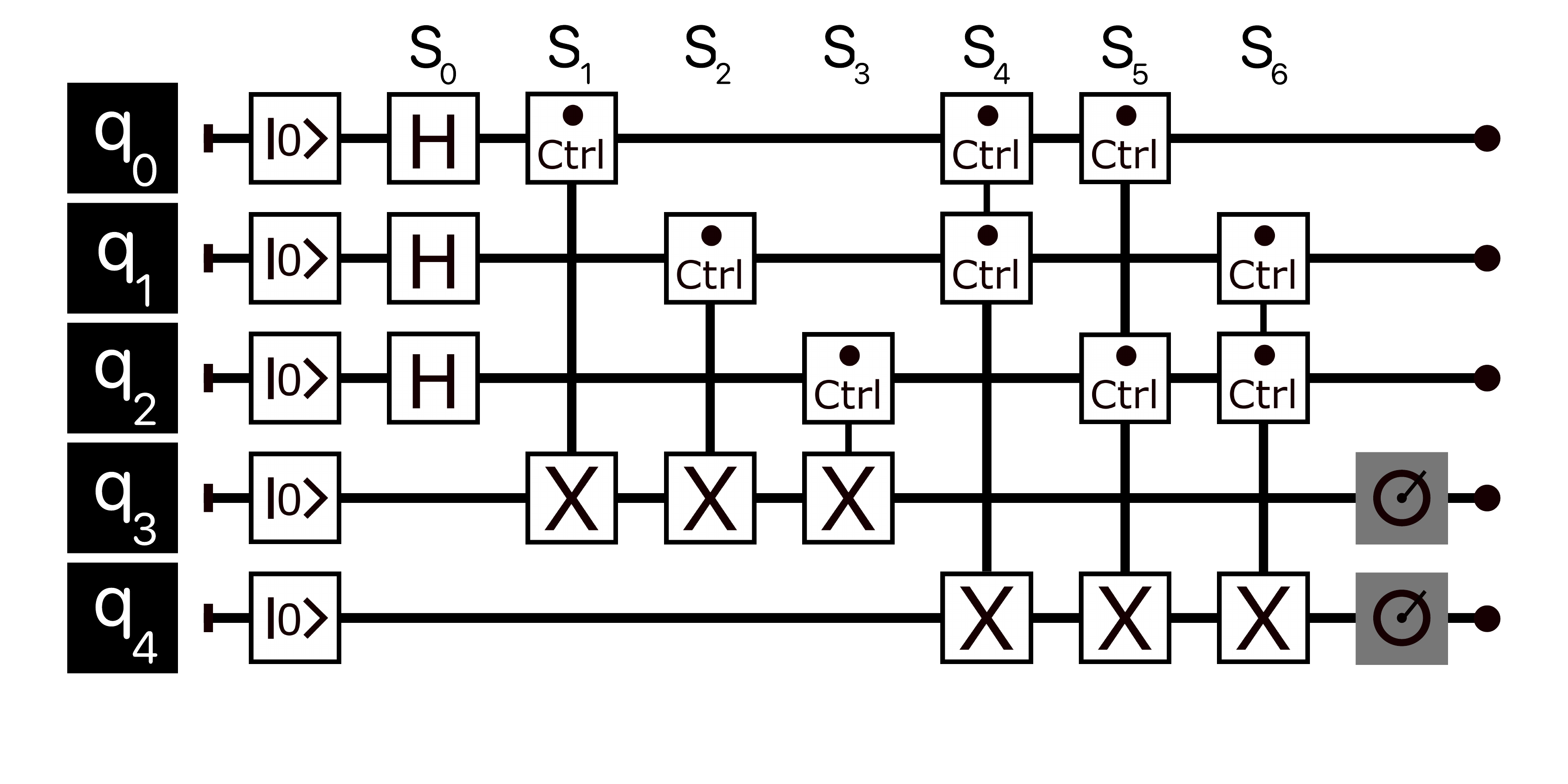}
\caption{Circuit diagram of a five qubits quantum full-adder in Psitrum. The first three qubits take the input and output bits are measured in the last two qubits.}
\label{F3}
\end{figure*}

Full-Adder circuit adds the input bits of $A$ and $B$ plus a carry input bit $C_{in}$ to produce the sum $S$ and a carry output $C_{out}$ bits. Classical full-adder requires three input and two output bits. However, the quantum version requires the same number of input and output qubits, since the circuit must be reversible. The truth table of the full-adder with assigned qubits is given in Table. \ref{FA}, and the corresponding quantum circuit is shown in Fig. \ref{F3}.

  \begin{table}[H]
  \begin{center}
  \begin{tabular}{ | c | c | c | c | c|}
  \hline
     $A$        &   $B$      &   $C_{in}$ &  $S$  & $C_{out}$  \\ \hline
     $q_{0}$    &   $q_{1}$  &   $q_{2}$ &  $q_{3}$  & $q_{4}$ \\ \hline
      0         &       0    &       0    &   0   &      0     \\ 
      0         &    0       &       1    &   1   &      0     \\ 
      0         &    1       &       0    &   1   &      0     \\  
      0         &    1       &       1    &   0   &      1     \\ 
      1         &    0       &       0    &   1   &      0     \\ 
      1         &    0       &       1    &   0   &      1     \\ 
      1         &    1       &       0    &   0   &      1     \\ 
      1         &    1       &       1    &   1   &      1     \\ \hline
      
  \end{tabular}
  \caption{Truth table of a full-adder circuit. A quantum full-adder requires at least five qubits with the same number of inputs and outputs so it becomes reversible. $q_{0}$ and $q_{1}$ represent the input bits with $q_{3}$ representing the input carry bit. The output sum and carry bits are represented by $q_{3}$ and $q_{4}$, respectively.}
  \label{FA}
  \end{center}
  \end{table}

\begin{figure*} [htp]
\includegraphics[width=1.0\textwidth]{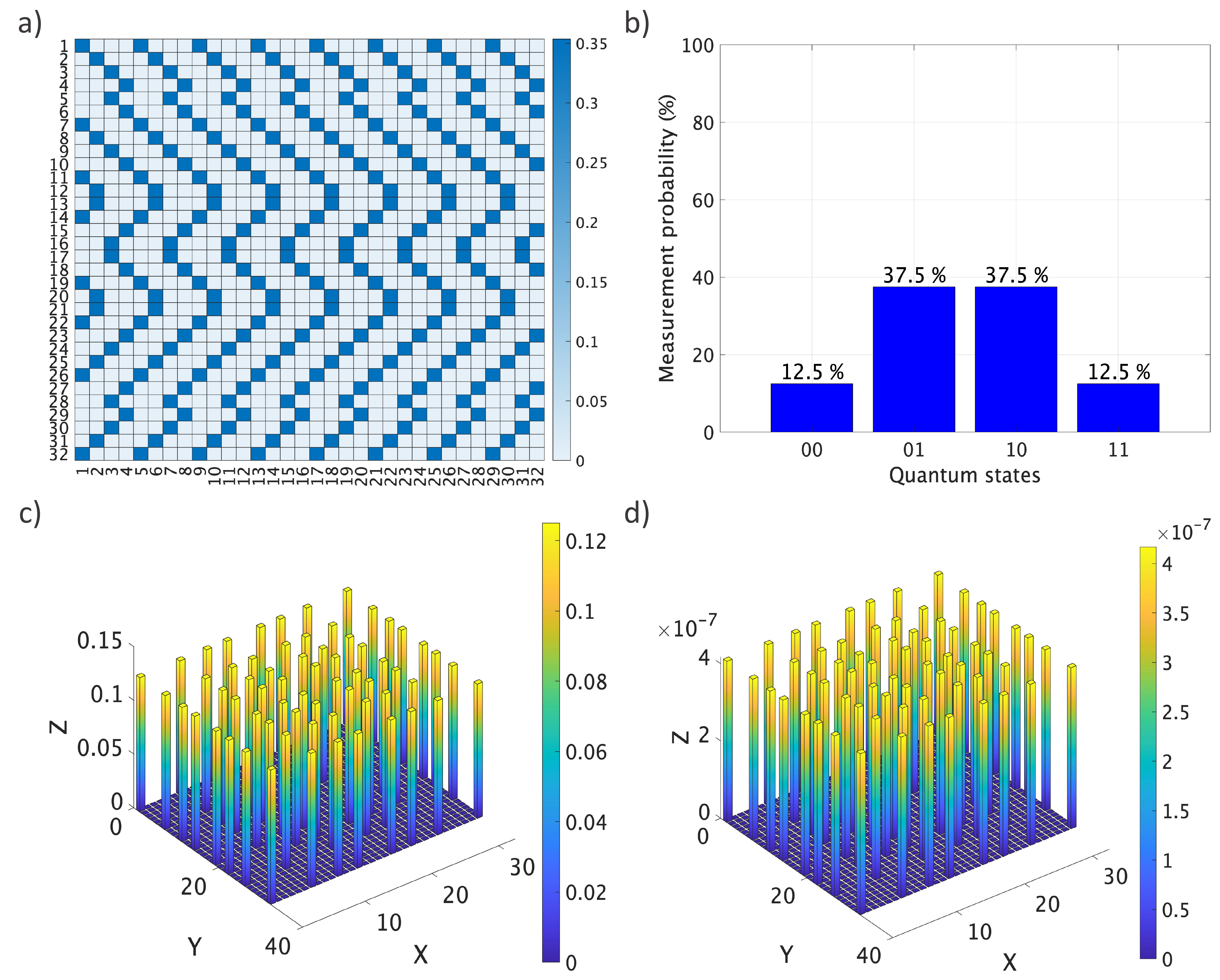}
\caption{Results after running quantum full-adder in Psitrum. (a) Shows heatmap of the simulated circuit matrix, and (b) shows the output states probabilities of measured qubits. (c) and (d) show the density matrix of the output with and without depolarizing noise channels, respectively. All stages are noisy depolarized with $P= 0.05$ at overshoot.}
\label{F4}
\end{figure*}

\pagebreak
Next, quantum full-adder is modeled in Psitrum while introducing depolarizing noise channels, given in equ. \ref{eq4}. Fig. \ref{F4}(a) shows a heatmap of the simulated circuit and Fig. \ref{F4}(b) shows the corresponding probabilities of the measured output qubits, matching the truth table in Table. \ref{FA}. Depolarizing channels are introduced at all stages of the circuit in Fig. \ref{F3}, with $P= 0.05$ at overshoot. The effect of this noise model can be seen on the output density matrix. Fig. \ref{F4}(c) and Fig. \ref{F4}(d) show the output density matrices with and without noise. Clearly, depolarizing channels reduce the amplitudes of the density matrix without introducing dephasing on qubits.  

\pagebreak
\subsection{Deutsch-Joza}
\begin{figure*} [htp]
\includegraphics[width=1.0\textwidth]{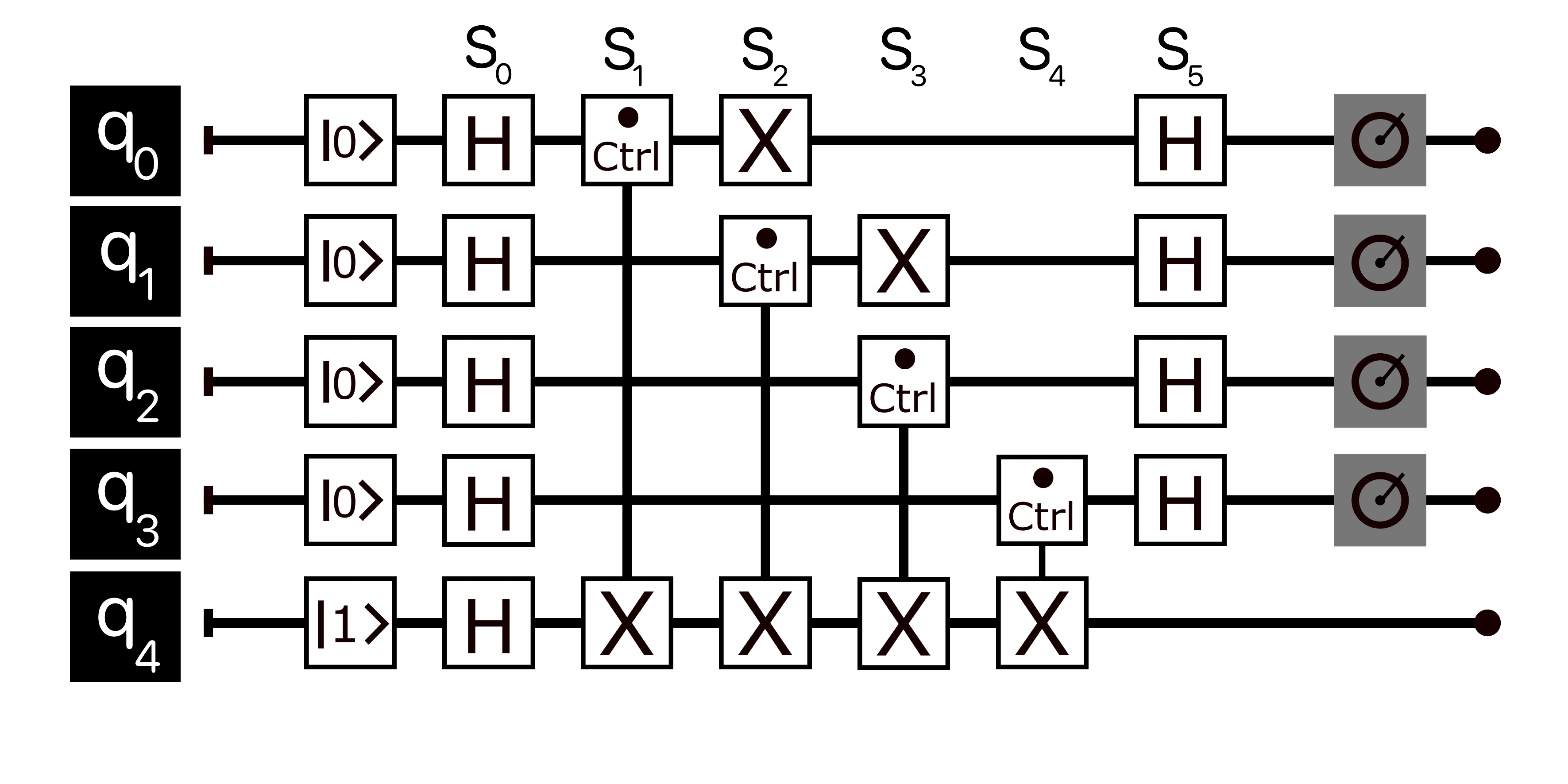}
\caption{Diagram of a Deutsch-Jozsa circuit of a balanced function in Psitrum. The first four qubits are measured to find out whether the oracle is balanced or constant.}
\label{F5}
\end{figure*}

Deutsch-Jozsa (DJ) algorithm finds whether an oracle is constant or balanced. If all outputs qubits are only zeros or ones then the function is constant, whereas if the function is balanced then exactly half the output qubits are measured to be zeros and the other half are ones. To apply DJ algorithm, first all qubits are initialized to zero states followed by Hadamard gates to create superposition. Next, the circuit of the oracle to be tested is constructed followed Hadamard gates on all qubits. Finally, the upper N-1 qubits are measured to find out whether this function constant or balanced. Fig. \ref{F5} shows DJ algorithm for a balanced function.
 
\begin{figure*} [htp]
\includegraphics[width=1.0\textwidth]{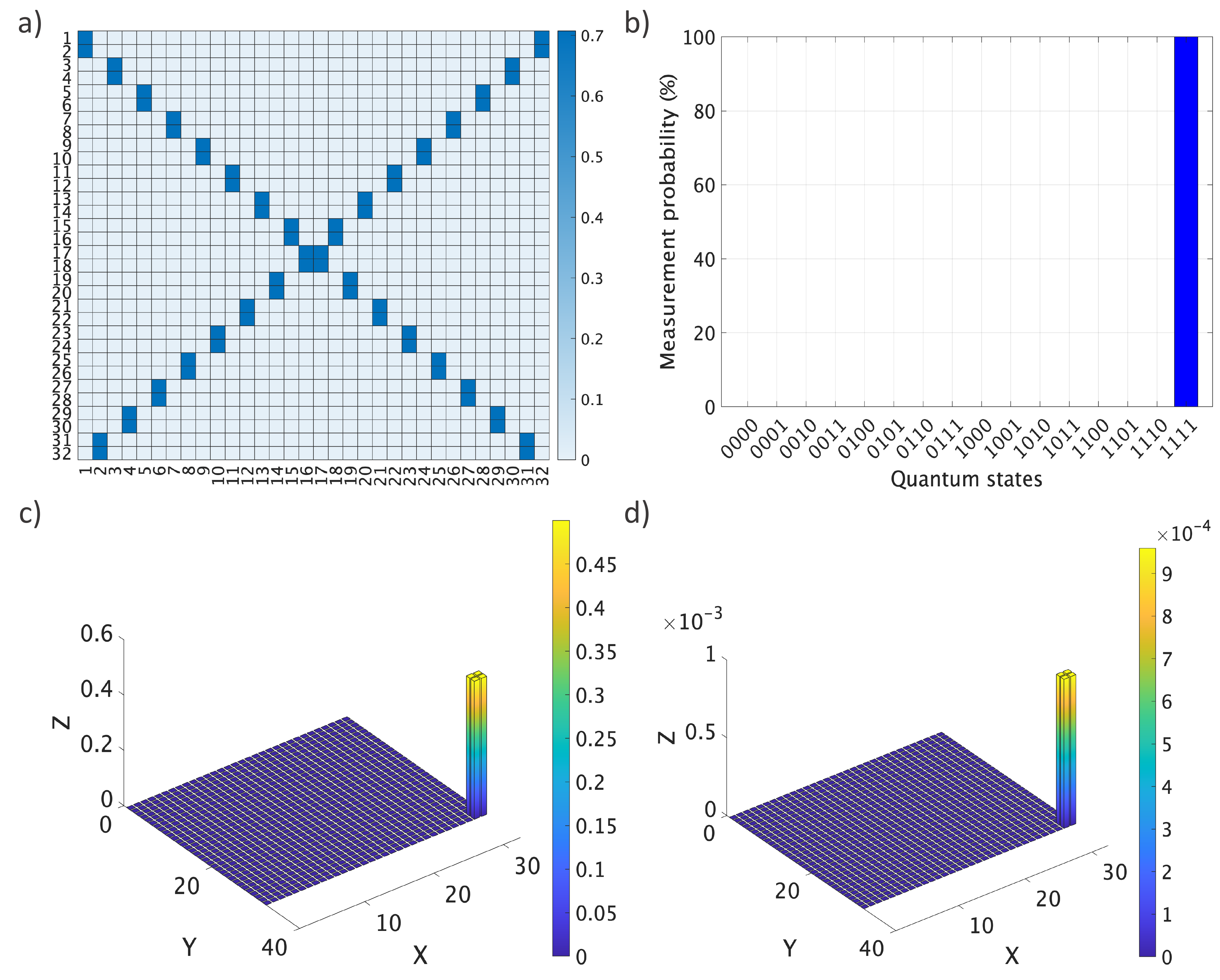}
\caption{Output of running Deutsch-Jozsa algorithm of a balanced function. (a) Shows heatmap of the simulated circuit matrix, and (b) shows the output states probabilities of measured qubits. (c) and (d) show the density matrix of the output with and without depolarizing noise channels, respectively. All stages are noisy depolarized with $P= 0.05$ at overshoot.}
\label{F6}
\end{figure*}

Next, DJ circuit is modeled in Psitrum with depolarizing noise channels. Fig. \ref{F6}(a) shows a heatmap of the simulated circuit and Fig. \ref{F6}(b) shows the measured probabilities of the first four qubits. All qubits are measured at the $\ket{1111}$ state, indicating that this is a balanced function. Depolarizing channels are introduced at all stages of the circuit in Fig. \ref{F5}, with $P= 0.05$ at overshoot. The effect of this noise model can be seen on the output density matrix. Fig. \ref{F6}(c) and Fig. \ref{F6}(d) show the output density matrices with and without noise. At both cases, the measured qubits are all still at the $\ket{1111}$ state with much smaller amplitudes. 

\pagebreak
\subsection{Grover search}
\begin{figure*} [htp]
\includegraphics[width=1.0\textwidth]{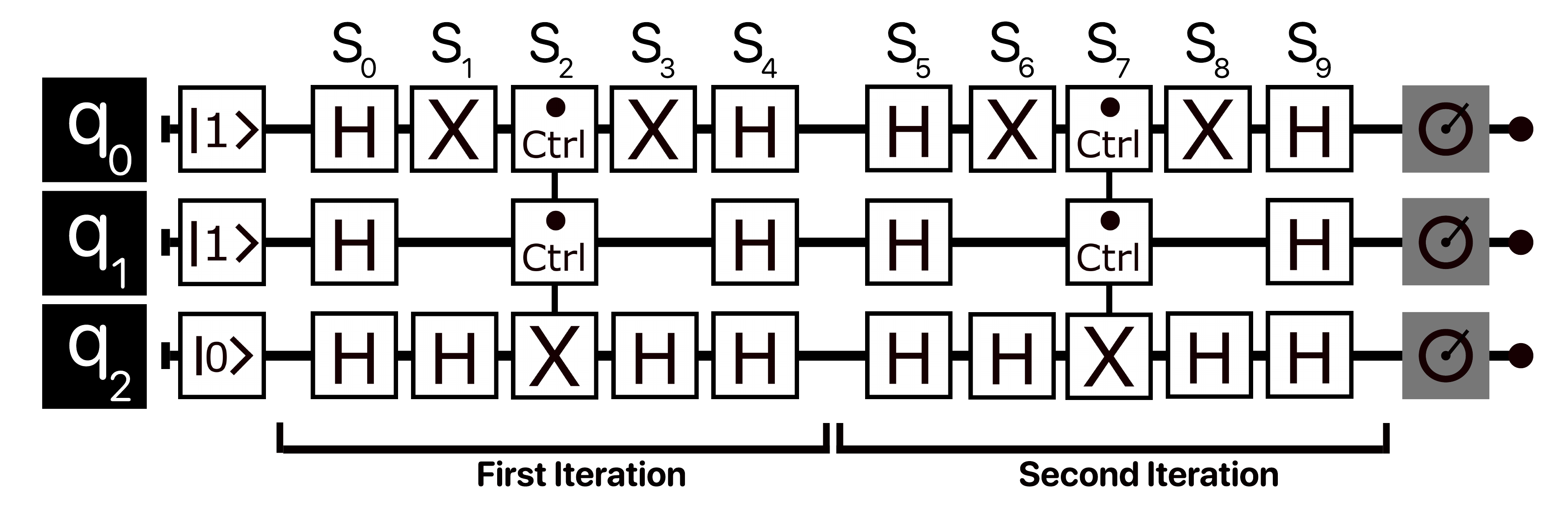}
\caption{Circuit diagram of two iterations of three qubits search algorithm. The qubit states are initialized at $\ket{110}$ representing the state to be searched.}
\label{F7}
\end{figure*}

Grover algorithm requires iterations that scale as the square root of length of the list. This is a quadratic speed over classical algorithms. First, qubits are initialized at the desired state to be found, followed by Hadamard gates to create superposition states. Next, Grover circuit applies selective phase inversion of the states followed by inversion about the mean in order to amplify the probability of measuring the correct state. Fig. \ref{F7} shows two iterations of Grover algorithm for three-qubit search of $\ket{110}$ state. 

\begin{figure*} [htp]
\includegraphics[width=1.0\textwidth]{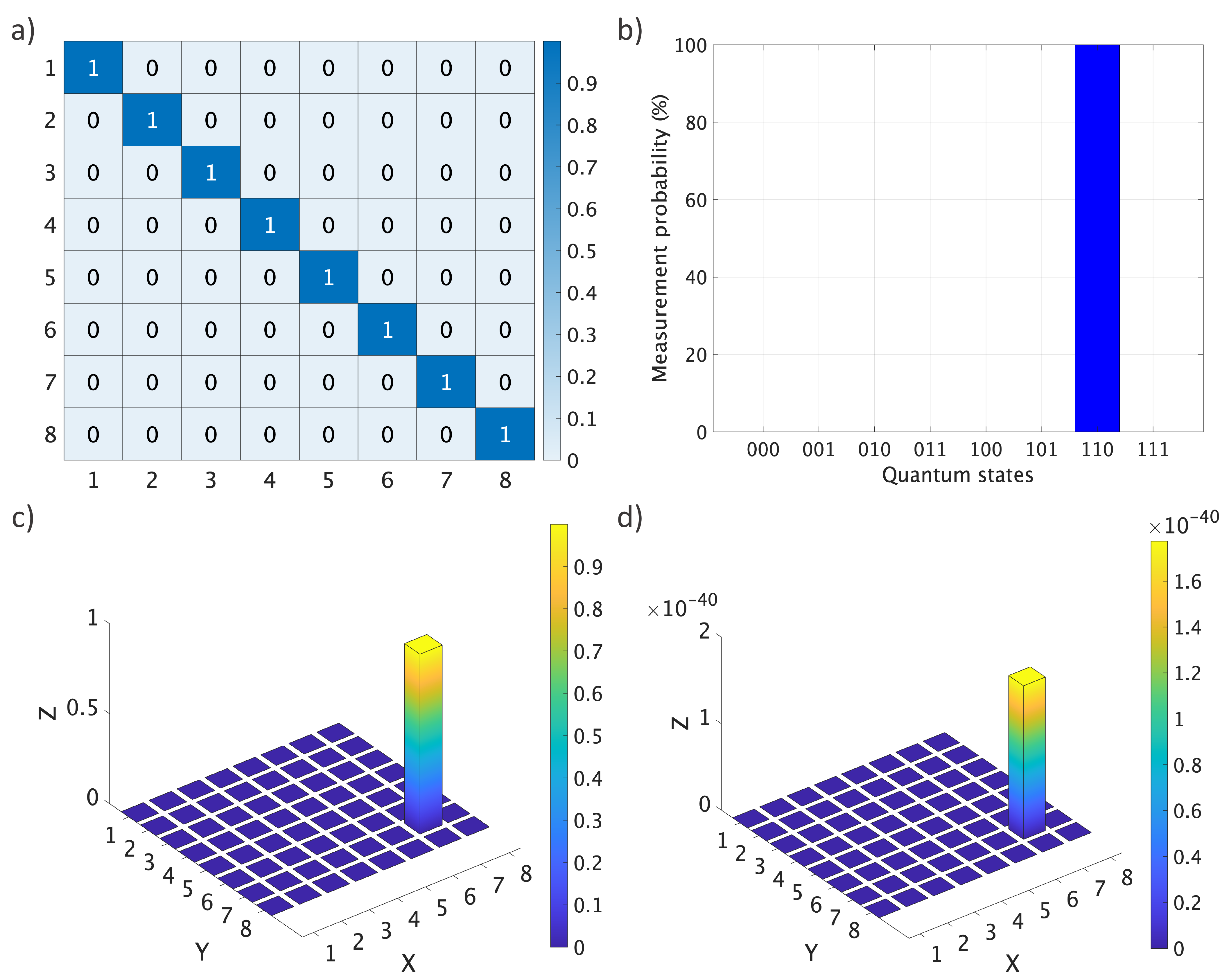}
\caption{Results of three qubits Grover algorithm. (a) Shows heatmap of the simulated circuit matrix, and (b) shows the output states probabilities of measured qubits after the second iteration. (c) and (d) show the density matrix of the output with and without depolarizing noise channels, respectively. All stages are noisy depolarized with $P= 0.05$ at overshoot.}
\label{F8}
\end{figure*}

Next, Grover circuit is modeled in Psitrum with  depolarizing noise channels. Fig. \ref{F8}(a) shows a heatmap of the simulated circuit and Fig. \ref{F8}(b) shows the measured probabilities of qubits after the second iteration. The correct answer is found in just two iterations, in which standard classical search would require at least eight iterations. Depolarizing channels are introduced at all stages of the circuit in Fig. \ref{F7}, with $P= 0.05$ at overshoot. The effect of this noise model can be seen on the output density matrix. Fig. \ref{F8}(c) and Fig. \ref{F8}(d) show the output density matrices with and without noise. The correct state still appears after depolarizing channels with smaller amplitudes.  

\pagebreak
\subsection{Prime factorization using Variational Quantum EigenSolver}
Unlike Shors algorithm that makes use of the period to compute the factors of the number to be factorized \cite{TV8}, here we compute factors of a given number through solving an optimization problem. The cost function used is given by $(N-pq)^2$, where $N$ is the number to be factorized, while $p$ and $q$ are factors to be identified and expressed in binary form as qubits over which the optimization is performed. Refer \cite{TV4} for a complete discussion on how the cost function is constructed and its complexity. The authors there made use of imaginary time evolution to solve for the factors. Here instead we only make use of the gradients to the cost function through standard Variational Quantum EigenSolver (VQE). We would like to note that the updates in the imaginary time evolution differs from VQE only by a factor of the fisher information that works similar to the hessian for gradient updates.

\begin{figure*} [htp]
\includegraphics[width=1.0\textwidth]{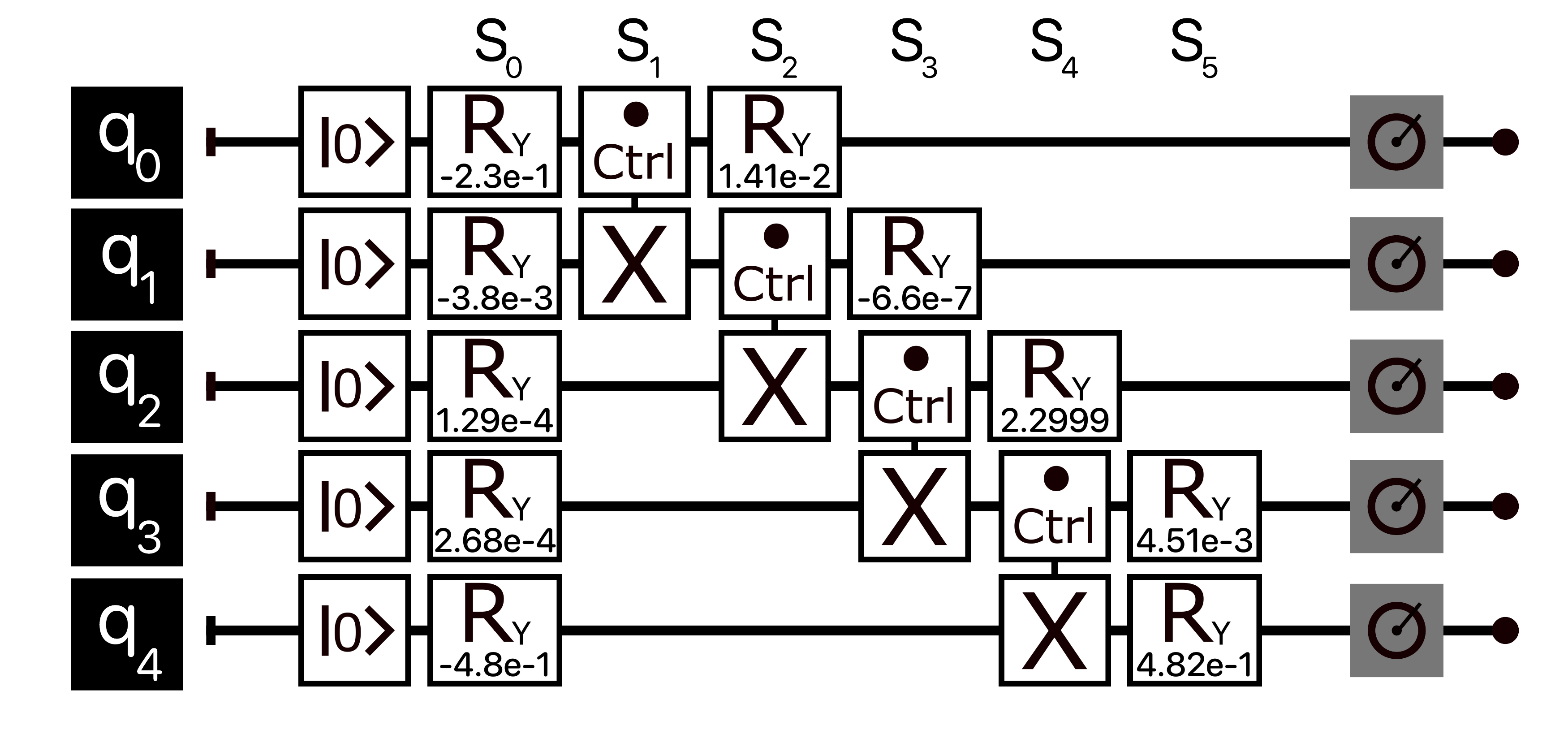}
\caption{Diagram of 5-qubit variational circuit used to factorize 91. Parameters of the circuits were randomly initialized.}
\label{F9}
\end{figure*}

\begin{figure*} [htp]
\includegraphics[width=0.95 \textwidth]{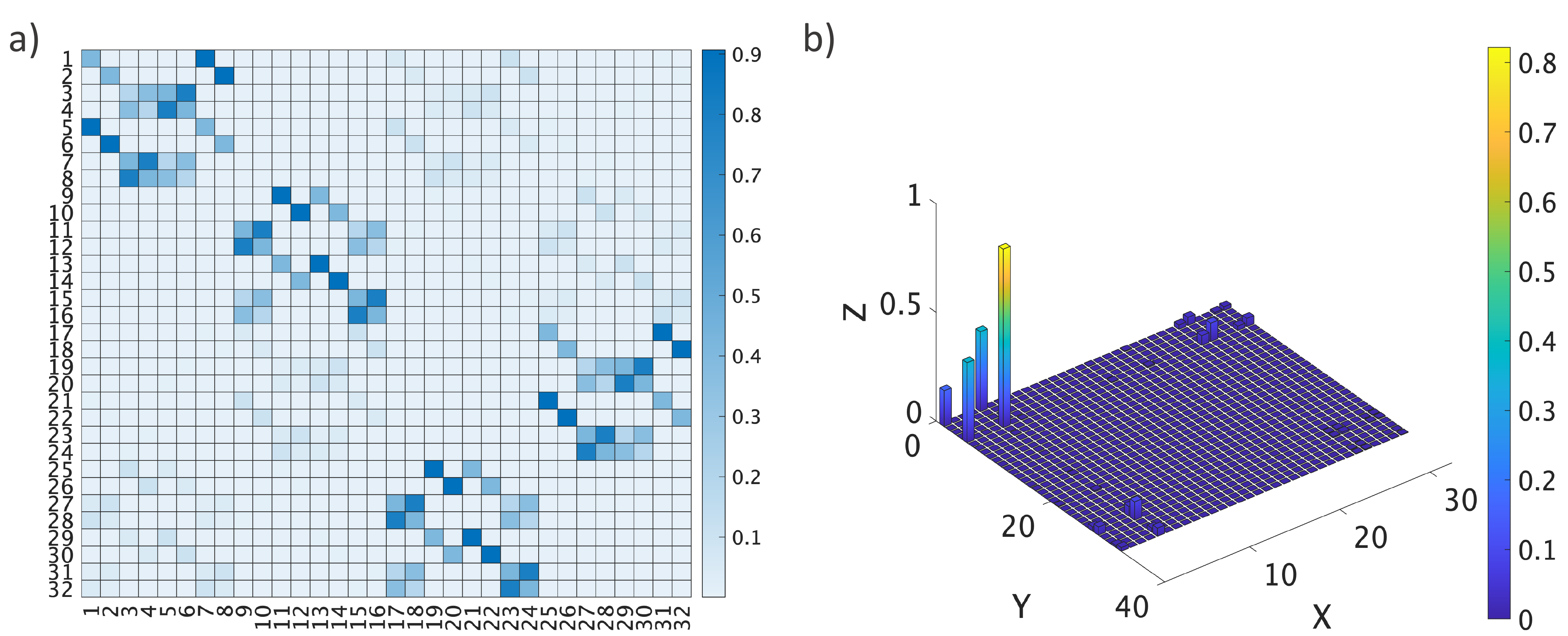}
\caption{Output of simulating the variational circuit for factoring 91 in Psitrum. (a) Shows heatmap of the circuit matrix, and (b) shows the density matrix of the output state.}
\label{F10}
\end{figure*}

We use a hardware efficient ansatz (Fig. \ref{F9}) to implement a variational circuit that consists of repeating layers to allow for a low error in the cost function output, but not representative enough to result in barren plateaus. The initial parameters of the circuit were randomly  initialized. The learning rate was set to $0.1$ and the convergence threshold for the amplitude of factors was set to $0.90$ at least. Results of simulating this variational circuit for factorizing 91 are shown in Fig. \ref{F10}.

\begin{figure*} [ht]
\includegraphics[width=1.0\textwidth]{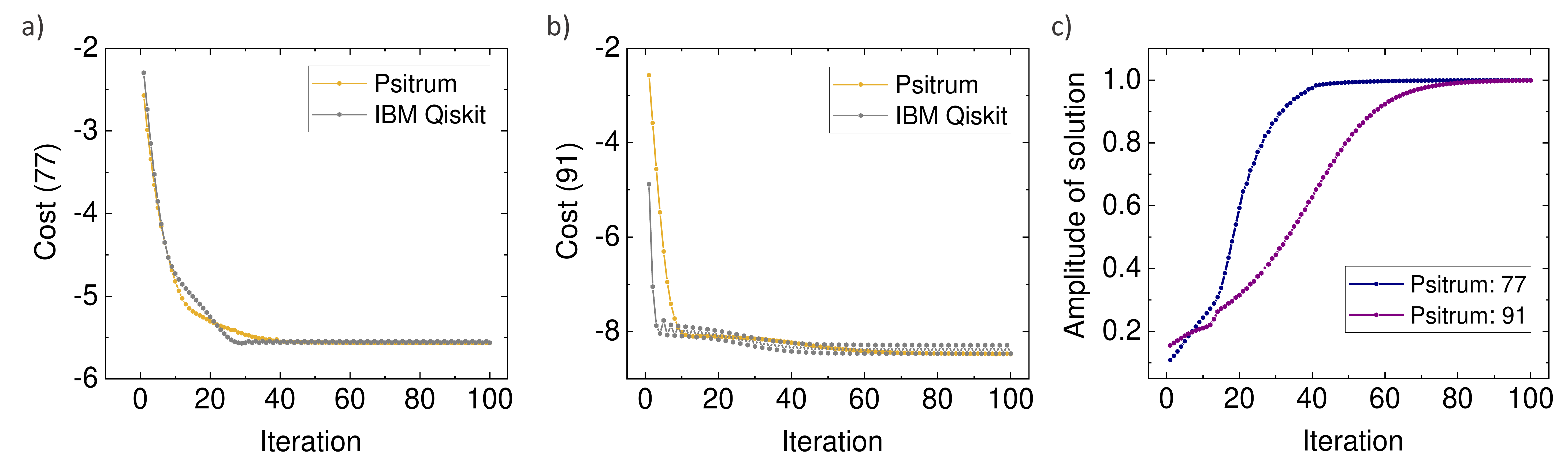}
\caption{5-qubit factorization examples. (a) and (b) show the minimization of the cost function for 77 and 91, respectively, using Psitrum and IBM Qiskit. (C) shows the amplitude of the solutions in the computational basis for factoring 77 and 91, respectively, as function of iteration solved by Psitrum.}
\label{F11}
\end{figure*}

\pagebreak
We plot training of the cost function and  the amplitude of the solutions at the end of $100$ iterations for the factorization of numbers $77$ and $91$ starting with the same set of initial parameters in either case and compare the results with IBM Qiskit Aqua framework (Fig. \ref{F11}). $77$ has factors of $11(1101)$ and $7(111)$, and $91$ has factors of $13(1110)$ and $7(111)$. Given that the least significant digit of all prime numbers begin with $1$, we have a total of 5 qubits over which the optimization is to be performed.

\pagebreak
\section{Conclusion}
We presented Psitrum, a universal quantum computer simulator based on classical hardware and showed how to run widely popular quantum algorithms, namely quantum-full adder, Deutsch-Joza and Grover search, both in the presence and absence of quantum gate noise. We made use of visualization tools available in the software to demonstrate the simulation of quantum circuits pertaining to these algorithms on Psitrum. In addition, we solved the factorization problem using standard Variational Quantum EigenSolver and were able to run the circuit in Psitrum to factorize prime numbers. This demonstrates that, Psitrum provides features for simulating noisy and noiseless quantum circuits to solve a wide class of quantum algorithms available for noisy intermediate-scale quantum processors. Given that Psitrum runs on a local server with a simplistic MATLAB interface, it provides a base layer for developers to easily add their own customization to suit their needs. In future, we are going to add modules relevant to implement various machine learning methods directly onto this platform, which will help to provide a good starting point to newcomers for a good experience with our visual learning tools.

\section{Acknowledgment}
This work was supported by the Deanship of Scientific Research at King Fahd University of Petroleum and Minerals. S.K. acknowledge the support from the National Science Foundation under award number 1955907.



\appendix 
\section{Quantum Gates in Psitrum}
\label{sec:sample:appendix}

This section provides details of gate operations implemented in Psitrum. These are the basic quantum gates for many algorithms and are available in this first version of Psitrum. In later versions, we can easily add more gates. Definitions of single- and multi-qubits gates are given in Table. \ref{AT1}. 

\pagebreak
      
     \newcolumntype{M}[1]{>{\centering\arraybackslash}m{#1}}
        \begin{longtable}{cM{50mm}M{50mm}M{50mm}}
           \toprule
           
            Gate & Matrix & Symbol  \\  
            \midrule
                     \centering \textbf{I} & $ \begin{bmatrix} 1 & 0 \\
                     0 & 1 \end{bmatrix} $  & 
                     \includegraphics[width=0.08\textwidth]{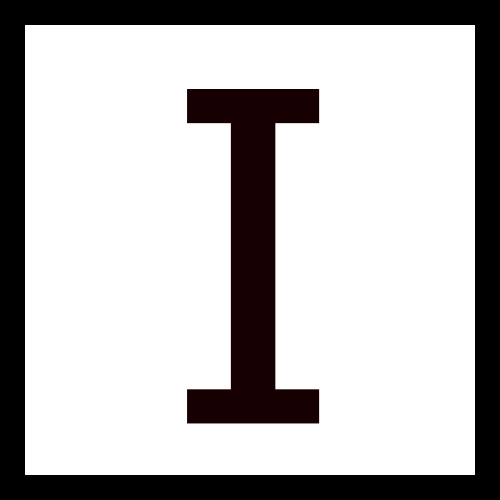}  \\ \\
                     
                     \centering \textbf{X} &  $ \begin{bmatrix} 0 & 1\\ 1 & 0 \end{bmatrix} $ & 
                     \includegraphics[width=0.08\textwidth]{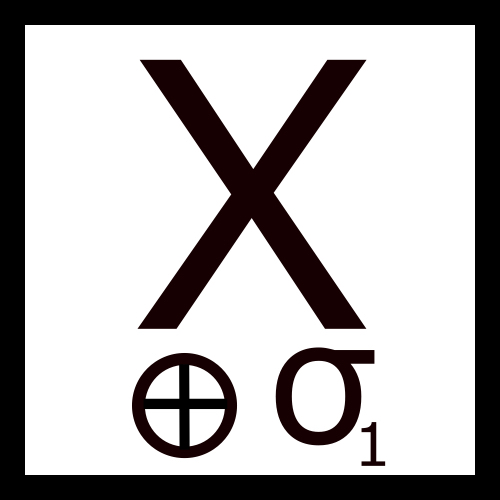} \\ \\
                     
                     \centering \textbf{Y} &  $ \begin{bmatrix} 0 & -i\\i & 0 \end{bmatrix} $ & 
                     \includegraphics[width=0.08\textwidth]{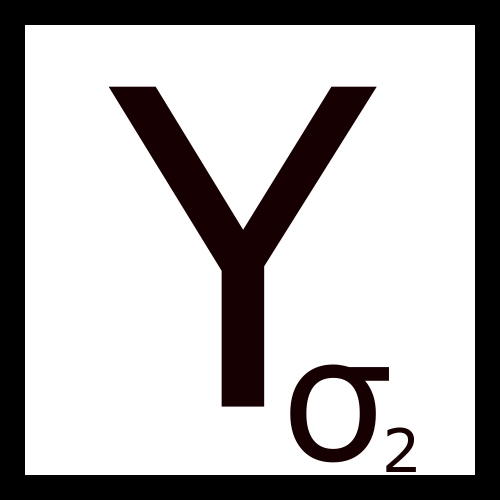} \\ \\
                     
                     \centering \textbf{Z} &  $ \begin{bmatrix} 1 & 0\\0 & -1 \end{bmatrix} $ & 
                     \includegraphics[width=0.08\textwidth]{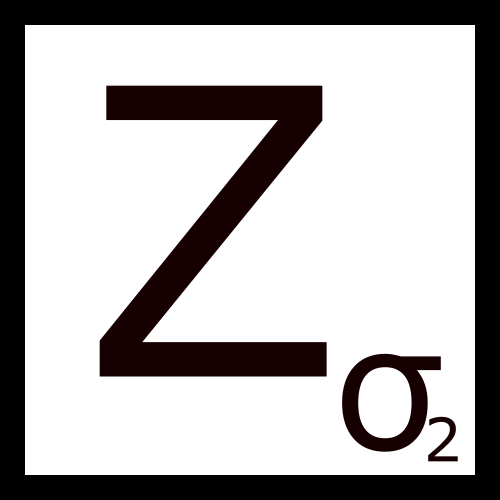} \\ \\
                     
                     \centering \textbf{H} &  $ \begin{bmatrix} 1 & 1\\1 & -1 \end{bmatrix} $ & 
                     \includegraphics[width=0.08\textwidth]{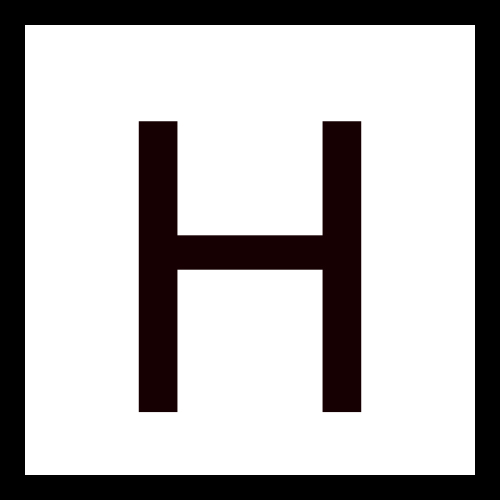} \\ \\
                     
                     \centering \textbf{S} &  $ \begin{bmatrix} 1 & 0\\0 & i \end{bmatrix} $ & 
                     \includegraphics[width=0.08\textwidth]{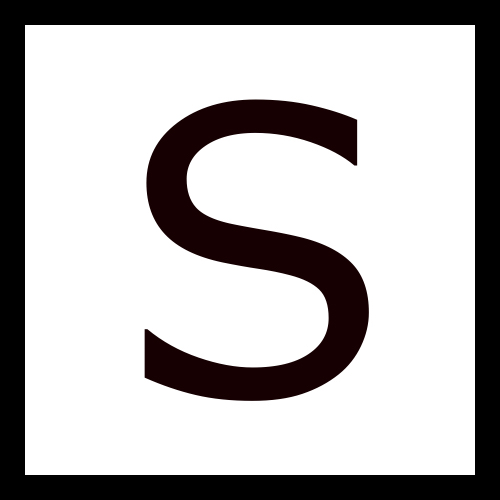} \\ \\
                     
                     \centering \textbf{T} &  $ \begin{bmatrix} 1 & 0\\0 & e^{\frac{i\pi}{4}} \end{bmatrix} $ & 
                     \includegraphics[width=0.08\textwidth]{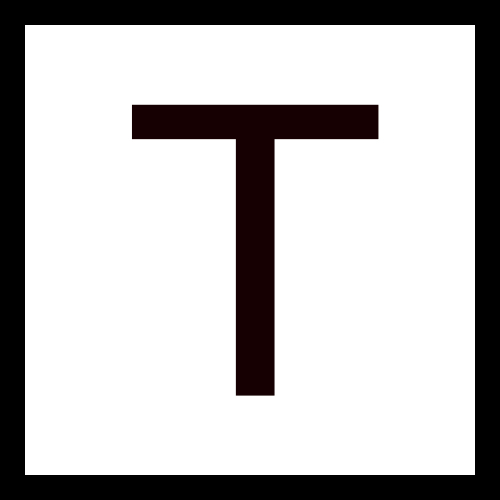} \\ \\
                     
                     \centering \textbf{$S^\dagger$} & $  \begin{bmatrix} 1 & 0\\0 & -i \end{bmatrix} $ &
                     \includegraphics[width=0.08\textwidth]{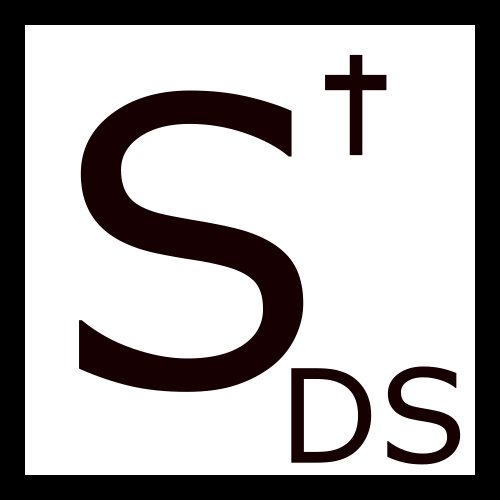} \\ \\
                     
                     \centering \textbf{$T^\dagger$} & $ \begin{bmatrix} 1 & 0\\0 & e^{-\frac{i\pi}{4}} \end{bmatrix} $ &
                     \includegraphics[width=0.08\textwidth]{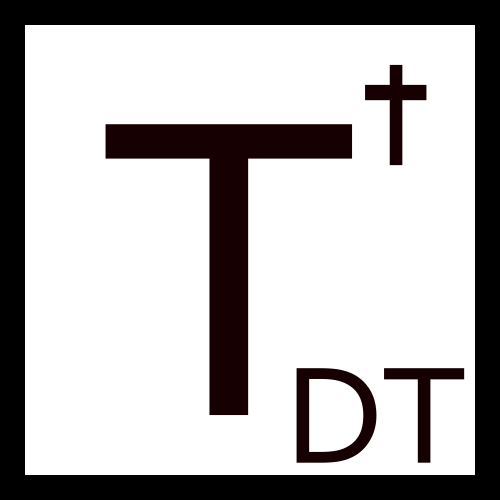} \\ \\
                     
                                         \centering \textbf{$U_{1}$} & $ \begin{bmatrix} 1 & 0\\0 & e^{i\theta} \end{bmatrix} $ &
                     \includegraphics[width=0.08\textwidth]{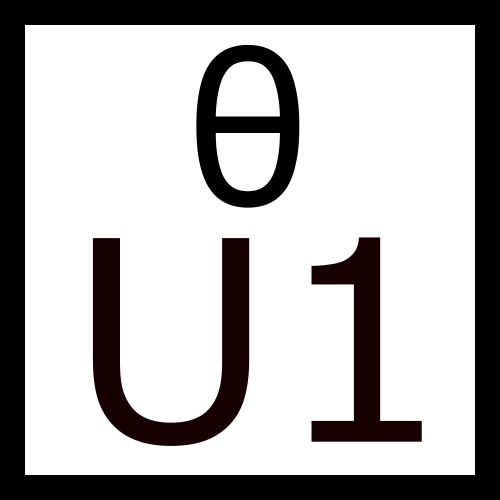} \\ \\
                     
                     \centering \textbf{$U_{2}$} &  $ \begin{bmatrix} 1 & -e^{i\theta}\\e^{i\phi} & e^{i(\theta+\phi)} \end{bmatrix} $ &
                     \includegraphics[width=0.08\textwidth]{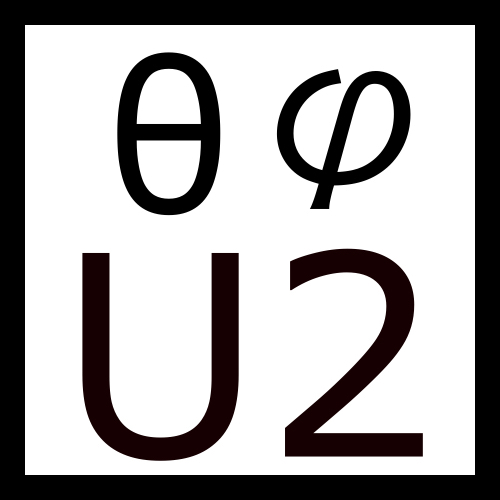} \\ \\
                     
                     \centering \textbf{$U_{3}$} &  $ \begin{bmatrix} \cos(\frac{\theta}{2}) & -e^{i\lambda}\sin(\frac{\theta}{2})\\e^{i\phi}\sin(\frac{\theta}{2}) & e^{i(\lambda+\theta)} \cos(\frac{\theta}{2}) \end{bmatrix} $ &
                     \includegraphics[width=0.08\textwidth]{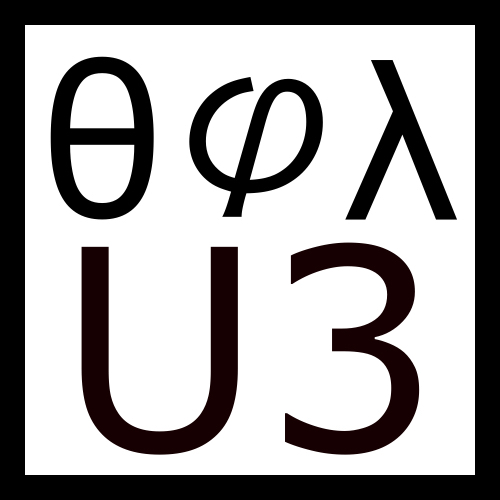} \\ \\
                     
                     \centering \textbf{$R_{X}$} & $ \begin{bmatrix} cos(\frac{\theta}{2}) & -isin(\frac{\theta}{2})\\-isin(\frac{\theta}{2}) & cos(\frac{\theta}{2}) \end{bmatrix} $ &
                     \includegraphics[width=0.08\textwidth]{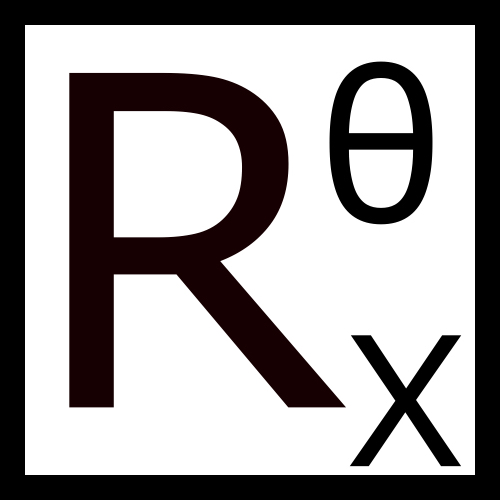} \\ \\
                     
                     \centering \textbf{$R_{Y}$} & $ \begin{bmatrix} cos(\frac{\theta}{2}) & -sin(\frac{\theta}{2})\\sin(\frac{\theta}{2}) & cos(\frac{\theta}{2})  \end{bmatrix} $ &
                     \includegraphics[width=0.08\textwidth]{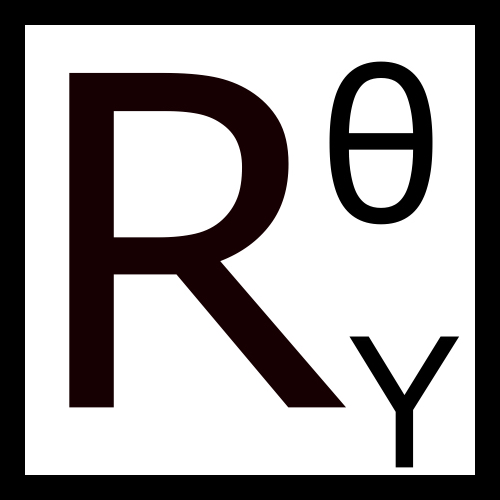} \\ \\
                     
                     \centering \textbf{$R_{Z}$} & $ \begin{bmatrix} e^{-\frac{i\phi}{2}} & 0\\0 & e^{\frac{i\phi}{2}}  \end{bmatrix} $ &
                     \includegraphics[width=0.08\textwidth]{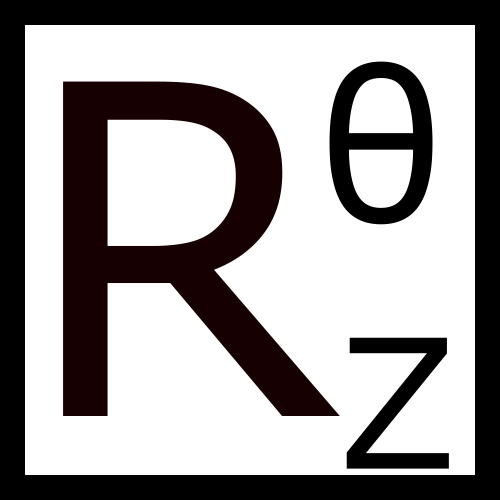} \\ \\ 
                     
                     \centering \textbf{$S_{X}$} & $ \begin{bmatrix} 1+i & 1-i\\1-i & 1+i  \end{bmatrix} $ &
                     \includegraphics[width=0.08\textwidth]{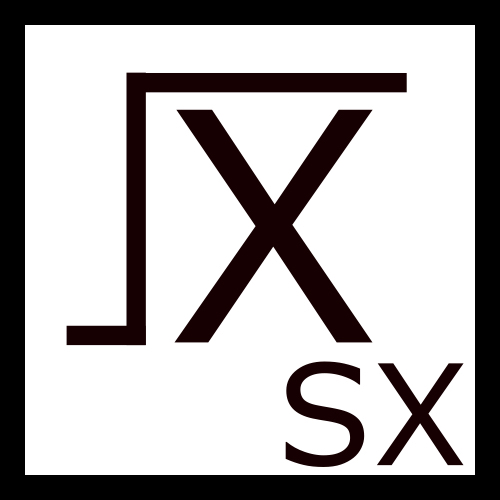} \\ \\
                     
                     \centering \textbf{$S_{X}^\dagger$} & $ \begin{bmatrix} 1-i & 1+i\\1+i & 1-i \end{bmatrix} $ &
                     \includegraphics[width=0.08\textwidth]{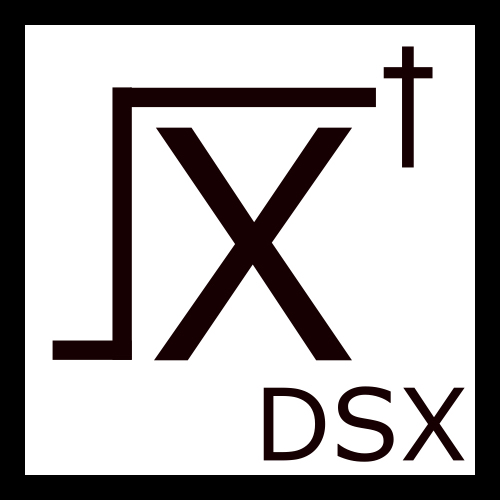} \\ \\
                     
                                          \centering \textbf{SWAP} & $ \begin{bmatrix}        1 & 0 & 0 & 0\\
                                                                        0 & 0 & 1 & 0\\
                                                                        0 & 1 & 0 & 0\\
                                                                        0 & 0 & 0 & 1 \end{bmatrix} $ &
                     \includegraphics[width=0.08\textwidth]{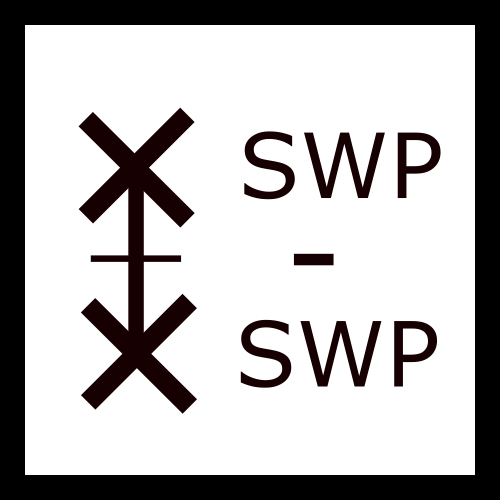} \\ \\
                     
                     \centering \textbf{CNOT} &  $ \begin{bmatrix}        1 & 0 & 0 & 0\\
                                                                        0 & 1 & 0 & 0\\
                                                                        0 & 0 & 0 & 1\\
                                                                        0 & 0 & 1 & 0 \end{bmatrix} $ &
                        \includegraphics[width=0.08\textwidth]{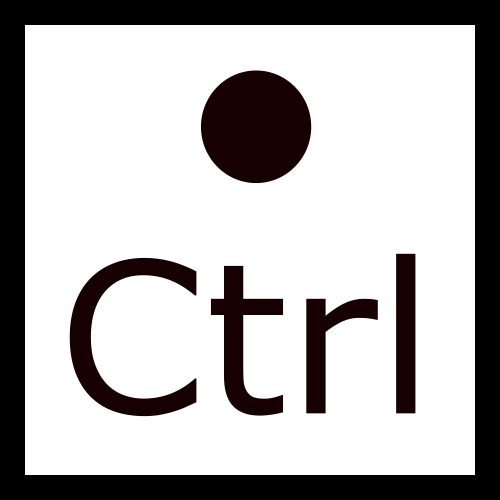} 
                        \includegraphics[width=0.08\textwidth]{X.jpeg}  \\ \\ 
                        
                     \centering \textbf{Toffoli} & $ \begin{bmatrix}
                                                        1 & 0 & 0 & 0 & 0 & 0 & 0 & 0\\
                                                        0 & 1 & 0 & 0 & 0 & 0 & 0 & 0\\
                                                        0 & 0 & 1 & 0 & 0 & 0 & 0 & 0\\
                                                        0 & 0 & 0 & 1 & 0 & 0 & 0 & 0\\
                                                        0 & 0 & 0 & 0 & 1 & 0 & 0 & 0\\
                                                        0 & 0 & 0 & 0 & 0 & 1 & 0 & 0\\
                                                        0 & 0 & 0 & 0 & 0 & 0 & 0 & 1\\
                                                        0 & 0 & 0 & 0 & 0 & 0 & 1 & 0 \end{bmatrix} $ &
                        \includegraphics[width=0.08\textwidth]{Ctrl.jpeg} 
                        \includegraphics[width=0.08\textwidth]{Ctrl.jpeg} 
                        \includegraphics[width=0.08\textwidth]{X.jpeg}  \\ \\  
                    
        \bottomrule
        \caption{Single- and multi-qubits quantum gates defined in Psitrum.}
        \label{AT1}
        \end{longtable}
        
 \bibliographystyle{elsarticle-num} 
 \bibliography{Biography}

\end{document}